\documentclass[aps,prl,superscriptaddress,twocolumn,showpacs,amssymb,amsmath]{revtex4}
\usepackage{graphics,epsfig,graphicx}
\usepackage{txfonts}
\usepackage[all]{xy}

\begin{document}

\title{Dynamic Behaviors in Directed Networks}
\author{Sung Min Park}
\affiliation{Center of Complex Systems, Samsung Economic Research Institute, 
Seoul 140-702, Korea}
\author{Beom Jun Kim}
\email[Corresponding author; ]{beomjun@skku.edu}
\affiliation{Department of physics and Institute of Basic Science, Sungkyunkwan University, Suwon
  440-746, Korea}

\begin{abstract}
Motivated by the abundance of directed synaptic couplings in a 
real biological neuronal network, we investigate the synchronization 
behavior of the Hodgkin-Huxley model in a directed network. 
We start from the standard model of the Watts-Strogatz undirected
network and then change undirected edges to directed arcs with a given 
probability, still preserving the connectivity of the network. 
A generalized clustering coefficient for directed networks
is defined and used to investigate the interplay between
the synchronization behavior and underlying structural
properties of directed networks. 
We observe that the directedness of complex networks plays
an important role in emerging dynamical behaviors, which
is also confirmed by a numerical study of the
sociological game theoretic voter  model  on directed networks. 
\end{abstract}


\maketitle

Research of complex systems which brings complex networks into focus
has been an intensive and successful area in physics and
other disciplines~\cite{ref:Basic,ref:netrep}.
Within the complex network research community,
dynamic behaviors in complex networks have been drawn much
attention. Among a variety of collective dynamic behaviors, 
synchronization is the one of the most popular research
topics~\cite{ref:syncbook,ref:strogatz,ref:synchreview}, and it has been shown that the connection topology of
complex networks greatly influences the degree of 
synchronization~\cite{ref:netrep,ref:Syn1}. Recently, the question of
how to increase the synchronizability via weighted and asymmetric 
couplings has been pursued intensively~\cite{ref:Syn2}. Especially,
it has been shown that the synchronization is enhanced  significantly
when the couplings from old to young vertices are more abundant
than from young to old, indicating the importance of 
the directedness of complex networks~\cite{ref:hwang}. 
As a good and real example of synchronization, the synchronous firing of 
action potential in neuronal networks has been studied~\cite{ref:BioPaperSyn}.
In~\cite{ref:HHexcit},
the Hodgkin-Huxley (HH) model~\cite{ref:HHmodel1} has been simulated
on the Watts-Strogatz network~\cite{ref:WS}, which
has revealed that the underlying small-world network structure enhances
synchronization substantially.
Different from real biological neuronal networks, where
synaptic couplings connecting two neurons are naturally directed,
it has been assumed that the synaptic couplings are undirected.
In contrast,  the neuronal network of the nematode 
\textit{Caenorhabditis elegans} (\textit{C. elegans}) has been analyzed
and 90\% of synaptic couplings have been shown to be 
directed~\cite{ref:Celegans}. In other
words, only 10\% of synaptic couplings are bidirectional.

In the present paper, we fully consider the directions of edges
and investigate in a systematic way how the directedness
of complex networks changes collective dynamical behaviors.
Both the HH model in biology and the sociological voter model 
are studied on directed model networks, and it is unanimously
found that as the more bidirectional edges are changed to unidirectional
ones the system as a whole exhibits a worse ordering behavior.
We believe that the conclusion should hold in a variety of different 
natural and social systems since the sparser bidirectional edges
are the less efficient the information flow becomes.

The directed small-world network in the present study is constructed
as follows: (1) The Watts-Strogatz undirected network is first built in
the same way as in~\cite{ref:WS}. In detail, 
starting from the locally connected network of one-dimensional lattice 
with the connection range $15$ corresponding to the average degree 
$\langle K \rangle = 30$, each edge is picked
and then randomly rewired with the probability $P$ to other randomly
chosen vertex. The size $N$ of the network is defined as the
total number of vertices and we use $N=400$ below.
(2) Each edge built above has both directions, 
one incoming and the other outgoing, and thus we substitute each undirected
edge as two arcs with opposite directions (we in this work call a directed
edge as an arc). Each pair of arcs is visited one by one, 
and then with the probability $\alpha$, the direction of randomly chosen
one arc in the given pair is reversed. It should be noted that the network resulting
from the above procedure is a directed network and double links
connecting the same two vertices are also allowed. We believe that
double edges exist not only in a biological neuronal network, 
where two  neurons can have two synaptic couplings, 
but also in a sociological network, 
where two individuals can have two different ways of information
transfer. In the viewpoint of dynamics in this work, 
the two arcs connecting the same two vertices are equivalent 
to one arc with double weights. Consequently, the resulting network from 
the above procedure is a directed network with double arcs allowed
(or double weights allowed), characterized by two parameters $P$ and $\alpha$. 
When $\alpha = 0$, the network is identical
to the WS undirected network, whereas for the other limiting case of
$\alpha = 1$, all edges in the network become directed.
By changing the parameter $\alpha$ from zero to unity, one can systematically 
change the density of directed edges, while still
preserving the average degree $\langle K \rangle$ (both the number
of arcs per vertex and the number of directly attached vertices per
vertex are conserved). 
It should be noted that if directions of arcs are not taken into
account the directed network at any $\alpha$ has the exactly the same 
structure as the undirected one. 

We first investigate structural properties of our directed network
and measure the characteristic path length and the clustering
coefficient. The former, denoted as $l$ and defined
by $l^{-1} \equiv \sum_{i\neq j}d^{-1}_{ij}/{N(N-1)}$ with the shortest
path length $d_{ij}$ connecting vertices $i$ and $j$, is computed
in the same way as for an undirected network~\cite{ref:path}.
The shortest paths should be computed with the direction
taken into account, and accordingly, $d_{ij} \neq d_{ji}$ 
for a directed network.

The clustering coefficient for an {\em undirected} network has usually 
been defined as~\cite{ref:WS} 
\begin{eqnarray}
C^{(0)} & \equiv &  \frac{1}{N}\sum_i C_i^{(0)}, \nonumber \\
C_i^{(0)} & \equiv & \frac{ 2 E_i}{k_i(k_i - 1)} ,
\end{eqnarray}
where $E_i$ is the number of edges in the set of neighbors of the vertex $i$.
If all neighbor vertices of $i$ (with the degree $k_i$) 
are connected to each other, $E_i =  k_i(k_i - 1)/2$ and thus
$C_i^{(0)} = 1$.

\begin{figure}
\includegraphics[width=0.45\textwidth]{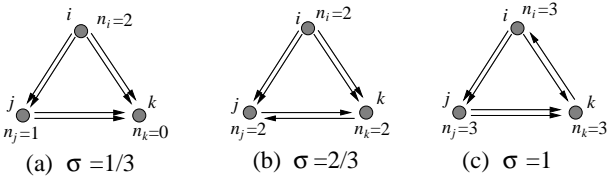}
\caption{Definition of the influence $\sigma$ of a triad. We define $n_i$ 
for the vertex $i$ as the number of vertices which
can get a message from $i$. A path connecting 
itself [e.g., $i \rightarrow j \rightarrow k \rightarrow i$ in (c)] is also counted.
In (a), $i$ can
send its message both to $j$ and $k$ and we assign
the number $n_i = 2$ for $i$; $n_j = 1$, $n_k = 0$ are
similarly assigned. In (b), $n_i = n_j = n_k = 2$, since,
e.g., $j$ can send message to $k$ and then can get it back from $k$.
In (c), $n_i = n_j = n_k = 3$ since everyone gets a message from
everyone. The influence $\sigma$ for a given triad $(i,j,k)$ 
is defined as $\sigma = (n_i + n_j + n_k)/9$. }
\label{fig:cc}
\end{figure}

For a directed network, however, the above definition of the clustering
coefficient needs to be changed. In this paper,  a quantity $\sigma_a$
(we call it the influence) is defined for the triad $a$ 
composed of three vertices
$a = (i,j,k)$ as follows: (1) Count the number $n_i$ of vertices (including
itself) who can get a message from $i$. 
For example, in Fig.~\ref{fig:cc}(a), 
$i$ can send its message to $j$ and $k$, and thus we assign $n_i = 2$.
In the same way, $n_j = 1$ and $n_k = 0$ are obtained in Fig.~\ref{fig:cc}(a). 
On the other hand, in Fig.~\ref{fig:cc}(b), $n_j = 2$ is assigned since
$j$ can send a message to $k$ and then get it back from $k$.
In Fig.~\ref{fig:cc}(c), $n_i = n_j = n_k = 3$ since a message from
anyone will be eventually delivered to everyone. (2) Since $n_i \leq 3$,
we normalize to obtain $\sigma_{a=(i,j,k)} \equiv (n_i + n_j + n_k)/9$
so that $0 \leq \sigma_{a=(i,j,k)} \leq 1$. In Fig.~\ref{fig:cc},
(a) $\sigma = (2+1+0)/9 = 1/3$, (b) $\sigma = (2+2+2)/9 = 2/3$, and
(c) $\sigma = (3+3+3)/9 = 1$, are found, respectively.
Although the examples of triad connections in Fig.~\ref{fig:cc}
have an identical number of arcs (six arcs), it is clear that the standard
definition of the clustering coefficient cannot capture the difference.
We define the clustering coefficient $C^{(d)}$ for a {\em directed} network as 
\begin{eqnarray}
\label{eq:directedcc}
C^{(d)} &\equiv& \frac{1}{N} \sum_i C_i^{(d)}, \nonumber \\
C_i^{(d)} &\equiv& \left(\frac{1}{E_i} \sum_{a=1}^{E_i} \sigma_a\right)
C_i^{(0)} = \frac{2}{k_i(k_i - 1)}\sum_{a=1}^{E_i} \sigma_a,
\end{eqnarray}
where $a = 1, 2, \cdots, E_i$ is the index for triads 
connected to the vertex $i$, and $\sigma_a$ is the 
influence defined above and shown in Fig.~\ref{fig:cc}.
Since $0 \leq \sigma_a \leq 1$, we get $\sum_a \sigma_a \leq E_i$
and consequently $C_i^{(d)} \leq C_i^{(0)}$ and $C^{(d)} \leq C^{(0)}$.

\begin{figure}
\includegraphics[width=0.23\textwidth]{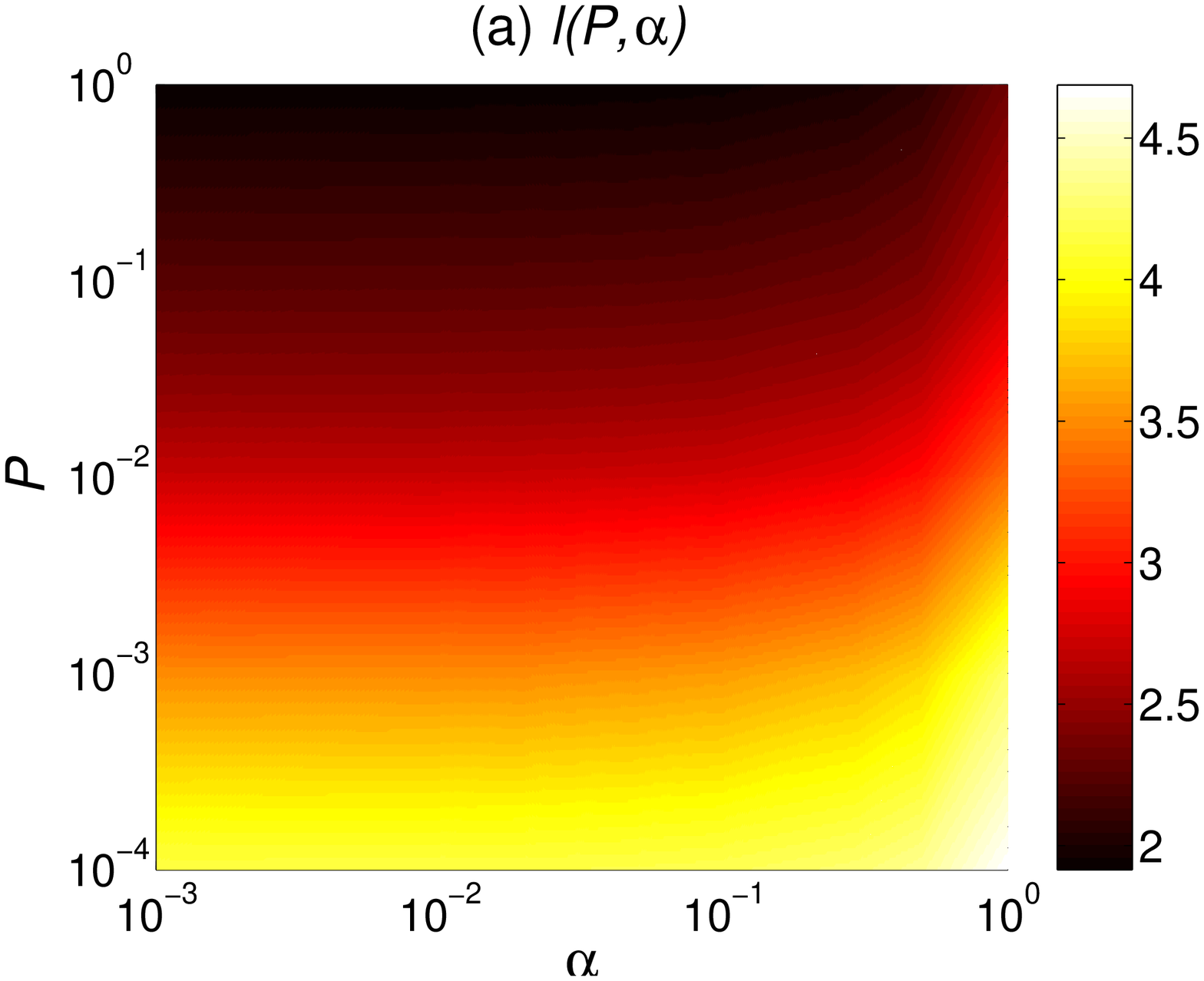}
\includegraphics[width=0.23\textwidth]{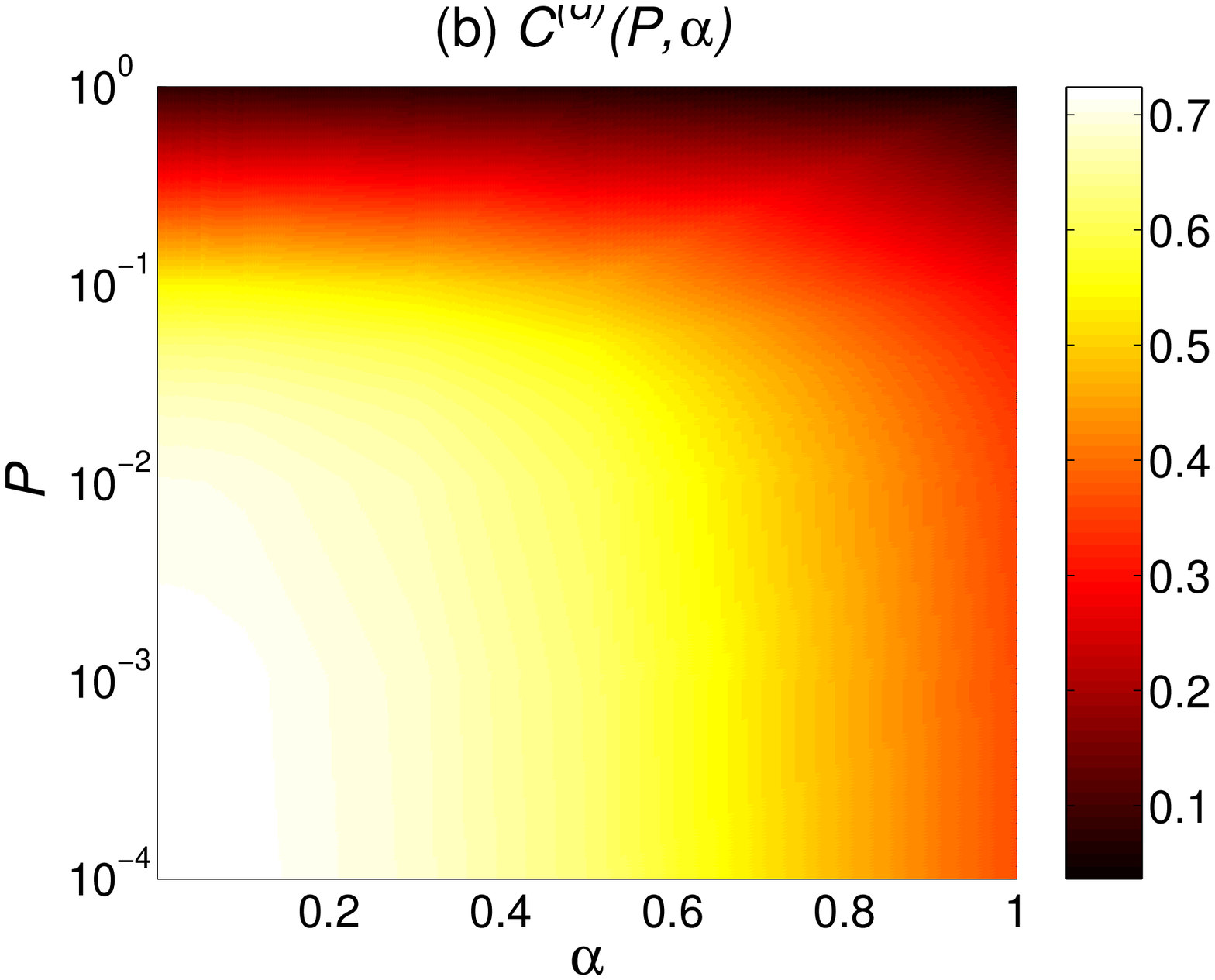}
\caption{(Color online) Density plots for (a) 
the characteristic path length $l$ 
and (b) the clustering coefficient $C^{(d)}$ in the plane of 
the rewiring probability $P$ and the probability $\alpha$ of
making directed arcs. Note that $P$ is in a log scale while $\alpha$ 
is not in (b).
As $P$ is increased, both  $l$ and $C^{(d)}$ are found to decrease
consistent with~\cite{ref:WS}. When $P =  10^{-2} \sim 10^{-1}$,
the network has short path length together with large degree
of clustering, manifesting the small-world behavior.
As more edges become directed arcs, i.e., as $\alpha$ is increased,
$l$ becomes larger while $C^{(d)}$ becomes smaller, weakening 
the  small-world behavior.
The network of the size $N=400$ at the average degree 
$\langle K \rangle =30$ has been used.}
\label{fig:structure}
\end{figure}

Figure~\ref{fig:structure} displays (a) $l(P,\alpha)$ and $C^{(d)}(P,\alpha)$
as density plots. At a fixed value of $\alpha$, the increase of $P$
results in smaller $l$ and $C^{(d)}$, consistent with the result
for the undirected network in~\cite{ref:WS} corresponding to $\alpha = 0$.  
The intermediate region of $P$ has short characteristic path length
but with a relatively high degree of clustering, manifesting the
so called small-world behavior~\cite{ref:WS}. In contrast, as 
the network has more directed arcs (i.e., as $\alpha$ is increased),
the small-world behavior becomes weaker: The path length is increased
and the clustering coefficient becomes smaller.
The structural changes related with the changing bidirectional
edges to unidirectional arcs are naturally expected to be
reflected in dynamic properties of the system on networks.
As more edges are changed to directed arcs, the path connecting
two vertices become longer, and the spread of information within
local neighbors become less efficient. 
We believe that our generalization of the clustering coefficient 
to capture the difference between directed and undirected networks 
can be very useful in similar studies. 
In Ref.~\cite{ref:triad}, a vector called the triad census has been proposed to
measure the frequencies of 16 isomorphism classes of all possible directed
coupling structures of three vertices. Among them seven correspond to complete
3-graphs like in Fig.~\ref{fig:cc}. Although it should be possible to use the
concept of the triad census~\cite{ref:triad} to characterize the clustering
property of directed networks, we believe that our suggestion of the clustering
coefficient is practically much more useful and convenient.

In order to study the effect of directedness of networks on dynamic
cooperative behaviors in detail,
we first study the Hodgkin-Huxley (HH) model in neuroscience 
on the directed network structure built as described above.
The HH model is the one of the most representative models 
describing dynamics of a neuronal system. Originally,
based on the result of physiological experiments of neural system 
of a squid, the HH model equations have been proposed to
describe the  membrane action 
potential~\cite{ref:HHmodel1,ref:ActionP1,ref:ActionP2}.
In the present work, we use the HH coupled differential 
equations in~\cite{ref:HHexcit} but with parameters
obtained from physiological experiments of
neurons in the part CA3 in hippocampus which plays an important
role in learning and memory~\cite{ref:HHmodel2}.
In detail, the membrane capacitance per unit area $C_m$=1.0$\mu$F/cm$^{2}$
and the external current density $I^{e}$=0.9$\mu$A/cm$^{2}$ are used, 
and maximum conductances for the leakage channel, 
the sodium, and the potassium ionic channels, are 
$g_L$=0.15$\Omega^{-1}/{\rm cm}^{2}$,  
$g_{\rm Na}$=50.0$\Omega^{-1}/{\rm cm}^{2}$, 
$g_{\rm K}$=10.0$\Omega^{-1}/{\rm cm}^{2}$, respectively
(see~\cite{ref:HHexcit} for HH coupled equations). 
The synaptic current between presynaptic and postsynaptic neurons
is treated in the same way as in~\cite{ref:HHexcit}.
In this work, a direct current is  injected into $40(=N/10)$ contiguous 
neurons  between the onset  (100msec) and the offset (1000msec)
times, and the 6th-order Runge-Kutta method with the time step
$\Delta t = 0.01$msec is used to integrate
HH equations. In order to get better statistics, 
500 independent runs are performed and then averaged over different
network realizations. 

We below use the synchronization order parameter in~\cite{ref:synchreview,ref:syncbook} 
to study the interplay between the dynamic synchronization behavior 
and the underlying directed network structure.  We denote $t_{i,n}$ as
the $n$th firing time of the neuron $i$, and define 
the corresponding phase variable $\theta_i(t)$ as 
\begin{equation}
\label{eq:theta}
\theta_{i}(t) \equiv  2\pi\frac{ t-t_{i,n-1} }{ t_{i,n}-t_{i,n-1} },
\end{equation}
via the simplification that 
$\theta(t)$ $(t_{i,n-1} \leq  t \leq t_{i,n})$  increases linearly 
from 0 to $2\pi$ between the two successive firings at 
$t_{i,n-1}$ and $t_{i,n}$. The synchronization order parameter 
is then defined  as
\begin{equation}
\label{eq:delta}
\Delta \equiv \frac{1}{N}\left\langle 
\sum_{i=1}^{ N_{\rm{fire}} }e^{i\theta_{i}(t)} \right\rangle ,
\end{equation}
where $\langle \cdots \rangle$ represents the time average 
after achieving a steady state, and $N_{\rm fire}$ is the number
of neurons which fired at least twice [otherwise $\theta(t)$ is
not defined]. 
For convenience, we have not
included winding number term in Eq.~(\ref{eq:theta}); it does
not change the value of $\Delta$. 

\begin{figure}
\includegraphics[width=0.23\textwidth]{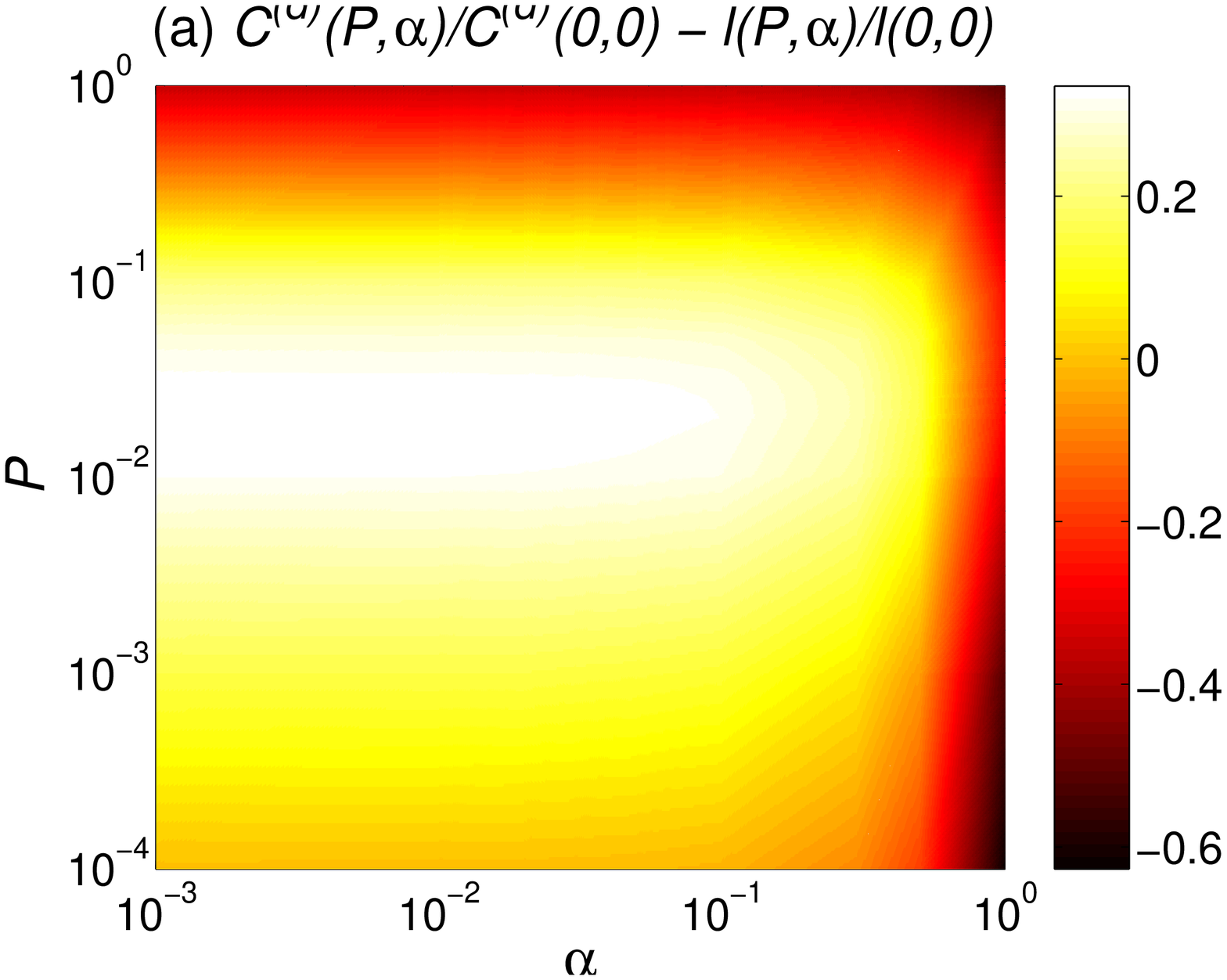}
\includegraphics[width=0.23\textwidth]{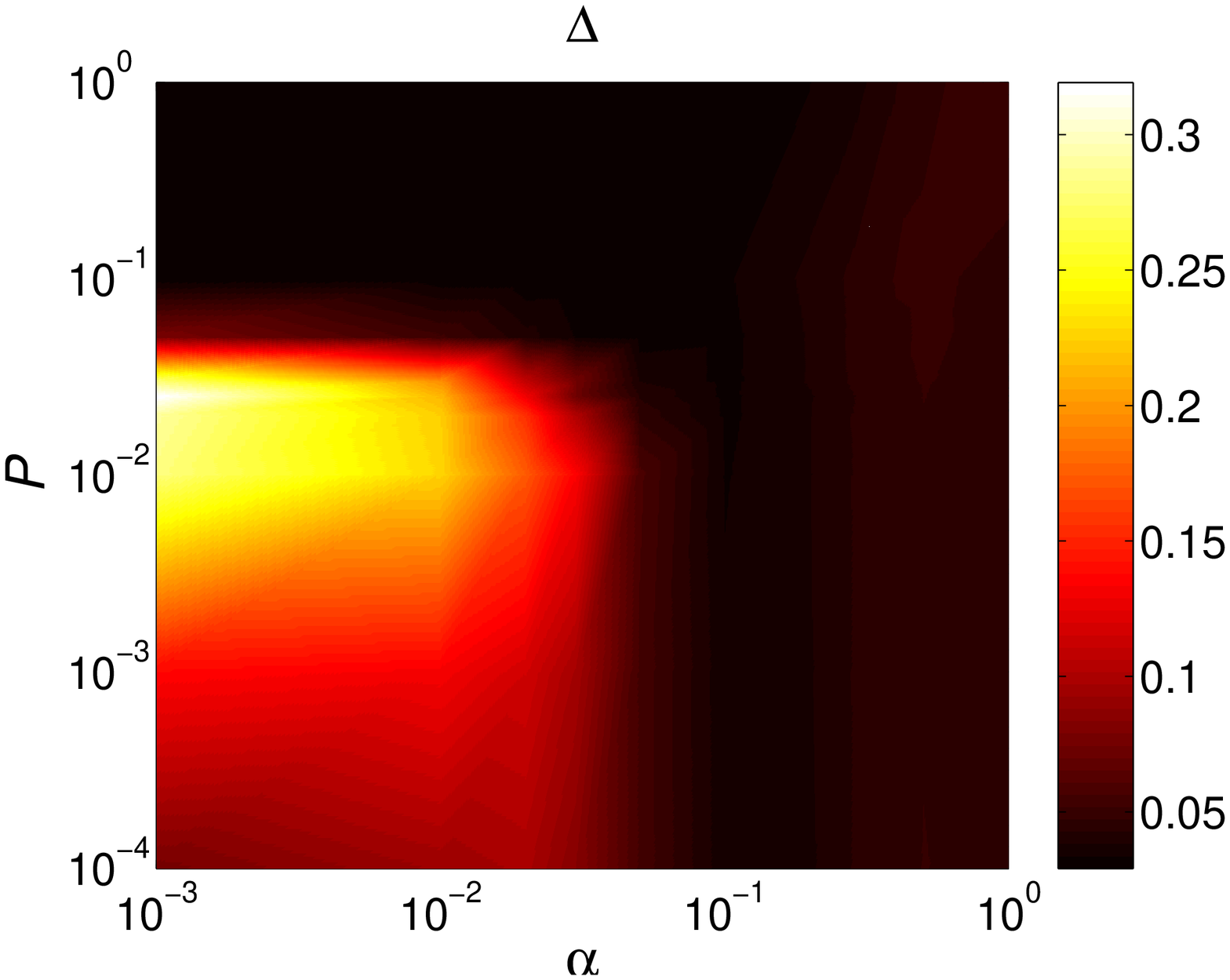}
\caption{(Color online) Density plots for (a) the small-world property
measured by $C^{(d)}(P,\alpha)/C^{(d)}(0,0)-l(P,\alpha)/l(0,0)$
and (b) the synchronization order parameter $\Delta(P,\alpha)$
computed for the Hodgkin-Huxley model in the $\alpha$-$P$ parameter space. 
As either $\alpha$ of $P$  becomes larger
the network loses both the structural small-world property 
and the dynamical synchronizability.  }
\label{fig:Delta}
\end{figure}

In Fig.~\ref{fig:Delta}, we compare (a) the structural small-world
property and (b) the dynamic synchronization behavior.
In order to measure the former in Fig.~\ref{fig:Delta}(a), we 
normalize the clustering coefficient and the characteristic path
length and then compute the difference, i.e., 
$\beta \equiv C^{(d)}(P,\alpha)/C^{(d)}(0,0) - l(P,\alpha)/l(0,0)$.
If $\beta$ has a sufficiently large value, the network has a relatively
large clustering coefficient and a relatively short path length, 
corresponding to the small-world region in~\cite{ref:WS}.
It is clearly displayed in Fig.~\ref{fig:Delta}(a), that as 
either $P$ or $\alpha$ is increased, the network begins to exhibit
the small-world property, which is destroyed eventually when $P$
or $\alpha$ is increased further. The structural importance of the 
directedness of networks is easily seen here: As more and more
edges are changed to directed arcs, the network loses the small-world
property. In Fig.~\ref{fig:Delta}(b), we show the synchronization 
order parameter $\Delta$ in Eq.~(\ref{eq:delta}) in the $\alpha$-$P$
plane. Although the region with large $\Delta$ values is much 
smaller than the region with the small-world behavior in 
Fig.~\ref{fig:Delta}(a), it is unambiguously shown that the existence
of directed arcs prohibits the system from being better synchronized.
We believe that if one neglects the directedness of networks 
in the study of structural and dynamic properties of networks,
the results should be taken cautiously. Especially, for a study
of neuronal networks, in which much more synaptic couplings
are directed than undirected, the conclusion reached by using
an undirected network may change significantly if one takes
the direction of synaptic couplings into full account. 

\begin{figure}
\centering{\resizebox*{0.40\textwidth}{!}{\includegraphics{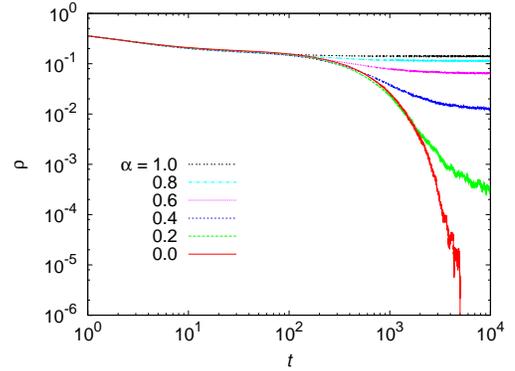}}}
\caption{(Color online) Fraction $\rho(t)$ of active bonds in the voter model
on directed networks as a function of time $t$.  From bottom to
top, curves correspond to $\alpha = 0.0$, 0.2, 0.4, 0.6, 0.8, and 1.0. 
As the network becomes more directed, i.e.,
as $\alpha$ is increased, the ordering towards the absorbing state
($\sigma_i = 1$ or $-1$ for all $i$) takes longer time, and the 
complete ordering becomes impossible for a sufficiently large 
$\alpha$ even at $t \rightarrow \infty$. Directed networks of 
the size $N=400$ with $\langle K \rangle = 30$ and the rewiring probability 
$P=0.1$ are used as an underlying interaction structure of the
voter model.
}
\label{fig:Voter}
\end{figure}

We next study a sociological game theoretic model called the
voter model~\cite{ref:voter}, defined on a directed network, 
to check the generality of our conclusion
drawn above for a biological system.
In the voter model, the $i$th player can have two different opinions
$\sigma_i = \pm 1$, and the time evolution of $\sigma_i (t)$ is
described as follows: (1) Pick one voter (call her $i$)  in  the
network randomly. (2) Pick another voter (call her $j$) randomly
among $i$'s incoming neighbors. (3) The voter $i$ changes her opinion
to the $j$'s one, i.e., $\sigma_i (t+1) = \sigma_j(t)$.
The directedness of a given network enters through the step (2) above:
Each voter is influenced only by her neighbor voters connected by 
{\it incoming} arcs. We use the random initial condition to start with,
and as time proceeds, the system approaches one of two absorbing states
characterized by a perfect ordering, i.e., $\sigma_i = 1$ or $\sigma_i = -1$
for all $i$.
In previous studies of voter models, the active bond is defined as the
one connecting two voters with different opinions, and 
the key quantity to measure is the fraction $\rho$ of active bonds given by
\begin{equation}
\rho=\frac{\sum_{i=1}^{N}\sum_{j\in \Lambda^{(in)}_{i}}(1-\sigma_{i}\sigma_{j})/2}{\sum_{i=1}^{N}K_{i}^{(in)}}, 
\end{equation}
where $\Lambda^{(in)}_i$ is the set of incoming neighbors of $i$,
and the incoming degree $K^{(in)}_i \equiv |\Lambda^{(in)}_i|$.
The active bond has different signs of the opinion $\sigma_i \sigma_j 
= -1$, and consequently $(1-\sigma_i \sigma_j)/2$ takes the value
either 1 or 0, depending on the activity of the bond.

In Fig.~\ref{fig:Voter}, the time evolution of $\rho$ is shown for
various values of $\alpha$ at the rewiring probability $P=0.1$
for the directed network of the size $N=400$ with the average
degree $\langle K \rangle = 30$.  It is displayed that as 
$\alpha$ is increased the convergence of opinions, i.e., 
$\rho \rightarrow  0$, takes longer time, and eventually becomes
impossible for large enough values of $\alpha$. 
This observation,  in parallel to the above finding  of the weaker
synchronizability of the HH model on directed networks, 
implies that the existence of directed arcs inhibits
an efficient flow of information, making global collective behaviors 
less plausible to occur.

In conclusion, we have studied dynamic behaviors
of the biological neuron model and the sociological voter model 
on directed networks, built from the standard network model
of Watts and Strogatz by changing undirected edges to directed arcs
with the probability $\alpha$. In the process of making
directed networks numbers of connected vertices and 
arcs are not changed, and the characteristic path length 
does not change significantly in a broad range of $\alpha$; 
this suggests that the connectivity is not altered much from
the undirected counterpart.
Unanimously found is that
as the network becomes more directed by the increment of $\alpha$,
a global emergence of collective behaviors (the synchronization 
in the former and the opinion convergence in the latter models)
becomes harder to develop. An extended definition of the clustering
coefficient for directed networks has also been suggested, and
the dynamic behaviors have been studied in relation 
with the network structures, captured by the characteristic
path length and the clustering coefficient.
However, we note that the region of the structural
small-world behavior in Fig.~\ref{fig:Delta} (a) and the region 
of the enhanced synchronizability in Fig.~\ref{fig:Delta} (b) 
do not overlap completely. The origin of this discrepancy
is not clear at present and needs further study in the future. 
We believe that our conclusion of decreased synchronizability
of directed networks is very general beyond the model systems 
studied in this work
and that the neglect of the directions of edges in network studies
needs to be done very carefully.

\acknowledgments 
B.J.K. was supported by grant No. R01-2005-000-10199-0
from the Basic Research Program of the Korea Science and Engineering
Foundation. S.M.P. acknowledges the support from Gu-Won research fund 
in Ajou University.

\end{document}